# AI virtues

## The missing link in putting AI ethics into practice

Dr. Thilo Hagendorff
thilo.hagendorff@uni-tuebingen.de
University of Tuebingen
Cluster of Excellence "Machine Learning: New Perspectives for Science"
International Center for Ethics in the Sciences and Humanities

**Abstract** – Several seminal ethics initiatives have stipulated sets of principles and standards for good technology development in the AI sector. However, widespread criticism has pointed out a lack of practical realization of these principles. Following that, AI ethics underwent a practical turn, but without deviating from the principled approach and the many shortcomings associated with it. This paper proposes a different approach. It defines four basic AI virtues, namely justice, honesty, responsibility and care, all of which represent specific motivational settings that constitute the very precondition for ethical decision making in the AI field. Moreover, it defines two second-order AI virtues, prudence and fortitude, that bolster achieving the basic virtues by helping with overcoming bounded ethicality or the many hidden psychological forces that impair ethical decision making and that are hitherto disregarded in AI ethics. Lastly, the paper describes measures for successfully cultivating the mentioned virtues in organizations dealing with AI research and development.

## 1   Introduction

Current AI ethics initiatives, especially when adopted in scientific institutes or companies, embrace a principle-based, deontological approach (Mittelstadt 2019). However, establishing principles alone does not suffice, they also must be convincingly put into practice. Most AI ethics guidelines do shy away from coming up with methods to accomplish this (Hagendorff 2020c). Nevertheless, recently more and more research papers appeared that describe steps on how to come "from what to how" (Morley et al. 2020; Vakkuri et al. 2019a; Eitel-Porter 2020; Theodorou and Dignum 2020). However, AI ethics still fails in certain regards. The reasons for that are manifold and reach from economical or legal to various other socio-cultural constraints. Economic imperatives in particular can overwrite ethical concerns and intentions. This is why both in academia and public debates, many authors state that AI ethics has not permeated the AI industry, quite the contrary (Vakkuri et al. 2019b). Despite the mentioned reasons, this is due to current AI ethics discourses hardly taking psychological considerations on bounded ethicality into account. They do not consider the limitations of the human mind, the many hidden psychological forces like powerful cognitive biases, bounded ethicality, blind spots and the like. In order to effectively improve ethical decision making in the AI field and to really live up to common ideals and expectations, AI ethics initiatives have to start choosing another path.

In order to effectively put AI ethics into practice and to deal with the mentioned constraints, one has to focus less on deontology, normative principles or rules. This is not to say that simply changing the sides to the great opponent of deontology, namely utilitarianism, is the solution. Utilitarianism can be of help when having to decide between two or more conflicting principles, for instance accuracy vs. fairness (Whittlestone et al. 2019). In this paper, though, another major ethical theory shall gain center stage. Instead of focusing on principles, AI ethics should put a stronger focus on virtues or, in other words, on character dispositions in AI practitioners in order to effectively put itself into practice. When using the term "AI practitioners" or "professionals", this includes AI or machine learning researchers, research project supervisors, data scientists, industry engineers and developers, as well as managers and other domain experts.

Moreover, to bridge the gap between existing AI ethics initiatives and the requirements for their successful implementation, one has to consider insights from moral psychology because, up to now, most parts of the AI ethics discourse disregard the psychological processes that limit the goals and effectiveness of ethics programs. This paper aims to respond to this gap in research. AI ethics, in order to be truly successful, has to stop repeating bullet points from the numerous ethics codes (Jobin et al. 2019). Instead, it should discuss the right dispositions in AI practitioners that can help not only to identify ethical issues and to engender the motivation to take action, but also – and this is even more important – to discover and circumvent one's own vulnerability to bounded ethicality, unconscious biases or other hidden psychological forces. The purpose of this paper is to state how this can be executed and how AI ethics can fix the missing link in order to actually put itself into practice.

## 2   AI ethics – The current principled approach

As mentioned in the introduction, current AI ethics programs come with several weaknesses and shortcomings. First and foremost, their normative principles lack reinforcement mechanisms (Rességuier and Rodrigues 2020). Basically, deviations from codes of ethics have no or very minor consequences. Moreover, even when AI applications fulfill all ethical requirements stipulated, it does not necessarily mean that the application itself is "ethically approved" when used in the wrong contexts or when developed by organizations that follow unethical intentions (Lauer 2020). In addition to that, ethics is often used for marketing purposes (Wagner 2018; Floridi 2019). Recent AI ethics initiatives of the private sector have faced a lot of criticism in this regard. In fact, industry efforts for ethical and fair AI are compared to past efforts of "Big Tobacco" to whitewash the image of smoking (Abdalla and Abdalla 2020). "Big Tech", so the argument, uses ethics initiatives and targeted research funds to avoid legislation or the creation of binding legal norms (Ochigame 2019). Hence, avoiding or addressing criticism like that is paramount for trustworthy ethics initiatives.

The latest progress in AI ethics research was configured by a "practical turn", which was among other things inspired by the conclusion that principles alone cannot guarantee ethical AI (Mittelstadt 2019). To accomplish that, so the argument, principles must be put into practice. Recently, several frameworks were developed, describing the process "from what to how" (Morley et al. 2020; Hallensleben et al. 2020; Zicari 2020). Ultimately, however, these frameworks are often just more detailed codes of ethics that use more fine-grained concepts than the initial high-level guidelines. For instance, instead of just stressing the importance of privacy, like the first generation of comprehensive AI ethics guidelines did, they hint to the Privacy by Design or Privacy Impact Assessment toolkits (Cavoukian et al. 2010; Cavoukian 2011; Oetzel and Spiekermann 2014). Or instead of just stipulating principles for AI, they differentiate between stages of algorithmic development, namely business and use-case development; design phase, where the business or use case is translated into tangible requirements for AI practitioners; training and test data procurement; building of the AI application; testing the application; deployment of the application; and monitoring of the application's performance (Morley et al. 2020). Other frameworks (Dignum 2018) are rougher and differentiate between ethics by design (integrating ethical decision routines in AI systems (Hagendorff 2020a)), ethics in design (finding development methods that support the evaluation of ethical implications of AI systems (Floridi et al. 2018)) and ethics for design (ensuring integrity on the side of developers (Johnson 2017)). But, as stated above, all frameworks still stick to the principled approach, even though they stress their practicality. The main transformation lies in the principles being far more nuanced and less abstract compared to the beginnings of AI ethics code initiatives (Future of Life Institute 2017). Typologies for every stage of the AI development pipeline are available. Differentiating principles solves one problem, namely the problem of too much abstraction. At the same time, however, it leaves many other problems open. Speaking more broadly, current AI ethics disregards many dimensions it should actually be having. In organizations of all kind, the likelihood of unethical decisions or behavior can be controlled to a certain extent. Antecedents for unethical behavior are individual characteristics (gender, cognitive moral development, idealism, job satisfaction, etc.), moral issue characteristics (the concentration and probability of negative effects, the magnitude of consequences, the proximity of the issue, etc.) and organizational environment characteristics (a benevolent ethical climate, ethical culture, code existence, rule enforcement, etc.) (Kish-Gephart et al. 2010). With regard to AI ethics, these factors are only partially considered. Most parts of the discourse are focused on discussing organizational environment characteristics (codes of ethics) or moral issues characteristics (AI safety) (Hagendorff 2020b, 2020c; Brundage et al. 2018), but not individual characteristics (character dispositions) increasing the likelihood of ethical decision making in AI research and development.

Therefore, a successful ethics strategy should focus on individual dispositions and organizational structures alike, whereas the overarching goal of every measure should be the prevention of harm. Or, in this case: prevent AI based applications from inflicting direct or indirect harm. This rationale might be fulfilled by ensuring explainability of algorithmic decision making, by mitigating biases and promoting fairness in machine learning, by fostering AI robustness and the like. However, more important than plainly listing these issues is asking how AI practitioners can be taught to intuitively keep them in mind. This would mean to transition from a situation of an external "ethics assessment" of existing AI products with a "checkbox guideline" to an internal process of establishing "ethics for design".

Empirical research shows that having plain knowledge on ethical topics or moral dilemmas is likely to have no measurable influence on decision making. Even ethics professionals, meaning ethics professors and other scholars of ethics, do not act more ethically than non-ethicists (Schwitzgebel and Rust 2014; Schwitzgebel 2009). Correspondingly, in the AI field, empirical research shows that ethical principles have no significant influence on technology developer's decision making routines (McNamara et al. 2018). Ultimately, ethical principles do not suffice to secure prosocial ways to develop and use new technologies (Mittelstadt 2019). Normative principles are not worth much if they are not acknowledged and adhered to. In order to actually acknowledge the importance of ethical considerations, certain character dispositions or virtues are required, among others, virtues encouraging to stick to moral ideals and values.

## 3 Basic AI virtues – The foundation for ethical decision making

What are the differences between principles and virtues? Basically, this differentiation is analogous to the difference between deontology and virtue ethics. While the former is based on normative rules that are universally valid, the latter addresses the question of what constitutes a good person or character. While ethical principles equal obligations, virtues are ideals that AI practitioners can aspire. Deontology focusses on the action rather than the actor. Thus, it defines action-guiding principles, whereas virtue ethics demands the development of specific positive character dispositions.

Why are these dispositions of importance for AI practitioners? One reason is that individuals, who display traits such as justice, honesty, empathy and the like, acquire (public) trust. Trust, in turn, makes it easier for people to cooperate and work together, it creates a sense of community and it makes social interactions more predictable (Schneier 2012). Acquiring and maintaining the trust of other players in the AI field, but also the trust of the general public, can be a prerequisite for providing AI products and services. After all, intrinsically motivated actions are more trustworthy in comparison to those which are simply the product of extrinsically motivated rule following behavior (Meara et al. 1996). Moreover, if AI practitioners have a reputation as virtuous persons, mistakes that will inevitably happen in the course of AI research and development will not have that much of a negative impact, but will rather be forgiven quickly.

One has to admit that a lot of ongoing AI basic research or very specific, small AI applications have such weak ethical implications that virtues or ethical values have no relevance at all. But AI applications that involve personal data, that are part of human-computer interaction or that are used on a grand scale, clearly have ethical implications that can be addressed by virtue ethics. In the theoretical process of transitioning from an "uncultivated" to a morally habituated state, "technomoral virtues" like civility, courage, humility, magnanimity and others can be fostered and acquired (Vallor 2016; Harris 2008; Kohen et al. 2019; Gambelin 2020; Sison et al. 2017; Neubert and Montañez 2020; Neubert 2017). In philosophy, virtue ethics traditionally comprises cardinal virtues, namely fortitude, justice, prudence and moderation. Nevertheless, in the context of AI applications, one has to sort out those virtues that are specifically important in the field of AI ethics. Based on patterns and regularities of the ongoing discussion on AI ethics, an ethics strategy that is based on virtues would constitute four basic AI virtues, where each virtue corresponds to a set of principles (see table 1). The basic AI virtues are justice, honesty, responsibility and care. But how exactly can these virtues be derived from AI ethics principles? Why do exactly these four virtues suffice?

When consulting meta-studies on AI ethics guidelines that stem from the sciences, industry, as well as governments (Jobin et al. 2019; Fjeld et al. 2020; Hagendorff 2020c), it becomes clear that AI ethics norms comprise a certain set of reoccurring principles. The mentioned meta-studies on AI ethics guidelines list these principles hierarchically, starting by the most frequently mentioned principles (fairness, transparency, accountability, etc.) and ending at principles that are mentioned rather seldom, but nevertheless repeatedly (sustainability, diversity, social cohesion etc.). When sifting through all these principles, one can, by using a reductionist approach and clustering them into groups, distill four basic virtues that cover all of them (see figure 1). The decisive question for the selection of the four basic AI virtues was: Does virtue *A* describe character dispositions that, when internalized by AI practitioners, will intrinsically motivate them to act in a way that "automatically" ensures or makes it more likely that the outcomes of their actions, among others, result in technological artefacts that meet the requirements that principle *X* specifies? Or, in short, does virtue *A* translate into behavior that is likely to result in an outcome that corresponds to the requirements of principle *X*? This question had to be applied for every principle that was derived from the meta-studies, testing by how many different virtues they can be covered. Ultimately, this process resulted in only four distinct virtues.

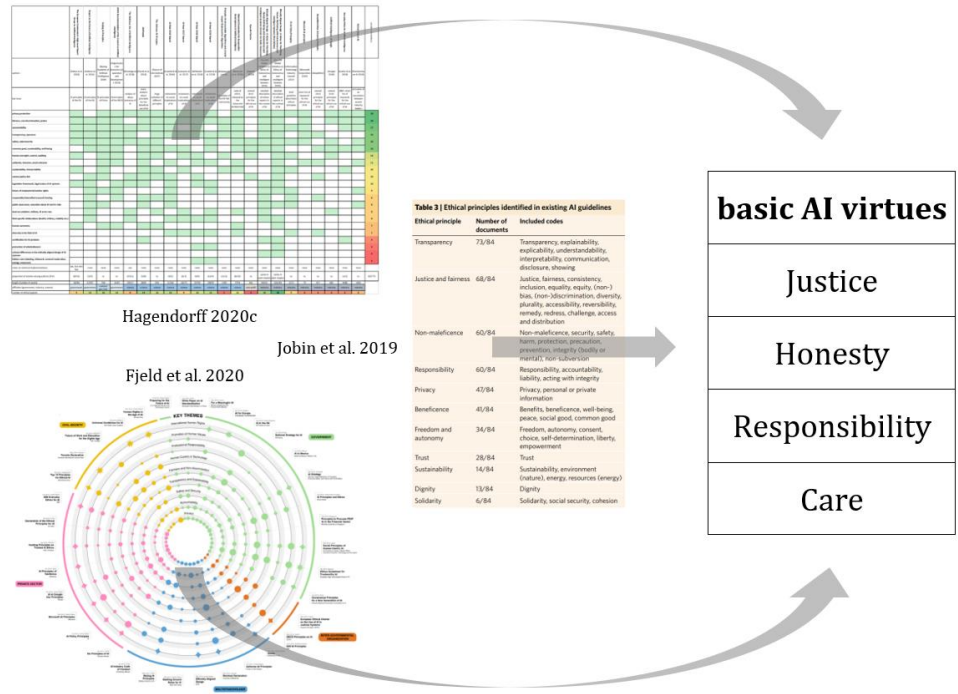

*Figure 1 - Using meta-studies on AI ethics guidelines as sources to distill four basic AI virtues*

To name some examples: The principle of algorithmic fairness corresponds to the virtue of justice. A just person will "automatically" be motivated to contribute to machine outputs that do not discriminate against groups of people, independently of external factors and guideline rules. The principle of transparency, as a second example, corresponds to the virtue honesty, because an honest person will "automatically" be inclined to be open about mistakes, to not hide technical shortcomings, to make research outcomes accessible and explainable. The principle of safe AI would be a third example. Here, the virtue of care will move professionals to act in a manner that they do not only acknowledge the importance of safety and harm avoidance, but also act accordingly. Ultimately, the transition happens between deontological rules, principles, or universal norms on the one hand and virtues, intrinsic motives, or character dispositions on the other hand. Nevertheless, both fields are connected by the same objective, namely to come up with trustworthy, human-centered, beneficial AI applications. Just the means to reach this objective are different. For the reasons mentioned, the paper argues that the means of virtue ethics can be more effective than the means of deontology when putting ethics into practice.

| Basic AI virtues | Explanation | Corresponding principles |
|---|---|---|
| **Justice** | A strong sense of justice enables individuals to act fairly, meaning that they refrain from having any prejudice or favoritism towards individuals based on their intrinsic or acquired traits in the context of decision making. In AI ethics, justice is the one moral value that seems to be prioritized the most. However, it is hitherto operationalized mainly in mathematical terms, not with regard to actual character dispositions of AI practitioners. Here, justice as a virtue could not just underpin motivations to develop fair machine learning algorithms, but also efforts to use AI techniques only in those societal contexts where it is fair to apply them. Eventually, justice affects algorithmic non-discrimination and bias mitigation in data sets as well as efforts to avoid social exclusion, fostering equality and ensuring diversity. | Algorithmic fairness, non-discrimination, bias mitigation, inclusion, equality, diversity |
| **Honesty** | Honesty is at the core of fulfilling a set of very important AI specific ethical issues. It fosters not only organizational transparency, meaning to provide information about financial or personnel related aspects regarding AI development. It also promotes the willingness to provide explainability or technical transparency regarding AI applications, for instance by disclosing origins of training data, quality checks the data were subject to, methods to find out how labels were defined etc. Moreover, honesty enables to acknowledge errors and mistakes that were made in AI research and development, allowing for collective learning processes. | Organizational transparency, openness, explainability, interpretability, technological disclosure, open source, acknowledge errors and mistakes |
| **Responsibility** | For the AI sector, responsibility is of great importance and stands in place of a host of other positive character dispositions. Mainly, responsibility builds the precondition for feeling accountable for AI technologies and their outcomes. This is particularly relevant since AI technology's inherent complexity leads to responsibility diffusions that exacerbate the assignment of wrongdoing. Diffusions of responsibility in complex technological as well as social networks can cause individuals to detach themselves from moral obligations, possibly leading to breeding grounds for | Responsibility, liability, accountability, replicability, legality, accuracy, considering (long term) technological consequences |
|  | unethical behavior. Responsibility, seen as a character disposition, is a counterweight to that since it leads professionals to actually feeling liable for what they are doing, opposing negative effects of a diffusion of responsibility. |  |
| **Care** | Care means to develop a sense for others' needs and the will to address them. Care has a strong connection to empathy, which is the precondition for taking the perspective of others and understanding their feelings and experiences. This way, care and empathy facilitate prosocial behavior and, on the other hand, discourage individuals from doing harm. In AI ethics, care builds the bedrock for motivating professionals to avoid AI applications from causing direct or indirect harm, ensuring safety, security, but also privacy preserving techniques. Moreover, care can motivate AI practitioners to design AI applications in a way that they foster sustainability, solidarity, social cohesion, common good, peace, freedom and the like. Care can be seen as being the driving force of the Beneficial AI movement. | Non-maleficence, harm, security, safety, privacy, protection, precaution, hidden costs, beneficence, well-being, sustainability, peace, common good, solidarity, social cohesion, freedom, autonomy, liberty, consent |

*Table 1 - List of basic AI virtues*

As said before, the four basic AI virtues cover all common principles of AI ethics as described in prior discourses (Floridi et al. 2018; Jobin et al. 2019; Morley et al. 2020; Hagendorff 2020c; Fjeld et al. 2020). They are the precondition for putting principles into practice by representing different motivational settings for steering decision making processes in AI research and development in the right direction. But stipulating those four basic AI virtues is not enough. Tackling ethics problems in practice also needs second-order virtues that enable professionals to deal with bounded ethicality.

## 4 Second-order AI virtues – A response to bounded ethicality

When using a simple ethical theory, one can assume that individuals go through three phases. First, individuals perceive that they are confronted with a moral decision they have to make. Secondly, they reflect on ethical principles and come up with a moral judgment. And finally, they act accordingly to these judgements and therefore act morally. But individuals do not actually behave this way. In fact, moral judgments are in most cases not influenced by moral reasoning (Haidt 2001). Moral judgements are done intuitively and moral reasoning is used in hindsight to justify one's initial reaction. In short, typically, moral action precedes moral judgment. This leads to consequences for AI ethics. It shows that parts of current ethics initiatives can be reduced to plain "justifications" for the status quo of technology development – or at least they are adopted to it. For instance, the most commonly stressed AI ethics principles are fairness, accountability, explainability, transparency, privacy and safety (Hagendorff 2020c). However, these are issues for which a lot of technical solutions already exist and where a lot of research is done anyhow. Hence,

AI ethics initiatives are simply reaffirming existing practices. On a macro level, this stands in correspondence with the aforementioned fact that moral judgments do not determine, but rather follow or explain prior decision making processes.

Despite explicit ethics training may improve AI practitioners' intellectual understanding of ethics itself, there are many limitations restricting ethical decision making in practice, no matter how comprising one's knowledge on ethical theories is. Many reasons for unethical behavior are resulting from environmental influences on human behavior and limitations through bounded rationality or, to be more precise, "bounded ethicality" (Bazerman and Tenbrunsel 2011; Tenbrunsel and Messick 2004). Hence, AI ethics programs are in need of specific virtues, namely virtues that help to "debias" ethical decision making in order to overcome bounded ethicality.

The first step to successively dissolve bounded ethicality is to inform AI practitioners not about the importance of machine biases, but psychological biases as well as situational forces. Here, two second-order virtues come into play, namely prudence and fortitude (see table 2). Whilst both virtues may help to overcome bounded ethicality, they are at the same time enablers for living up to the basic virtues. Individual psychological biases as well as situational forces can get in the way of acting justly, honestly, responsibly or caringly. Prudence and fortitude are the answers to the many forces that may restrict basic AI virtues, where prudence is aiming primarily at individual factors, while fortitude addresses supra-individual issues that can impair ethical decision making in AI research and development.

| **AI virtues** | **Explanation** | **Bounded ethicality** |
| --- | --- | --- |
| **Prudence** | Prudence means practical wisdom. In some philosophical theories, it represents the ability to gauge and reconcile complex and often competing values and requirements. Here, it stands for a high degree of self-understanding, for the ability to identify effects of bounded ethicality on one's own behavior as well as for the sincerity to acknowledge one's own vulnerability to unconscious cognitive biases. Prudence is the counterweight to the common limitations of the human mind, to the hidden psychological forces that impair ethical reasoning and decision making. | System 1 thinking, implicit biases, in-group favoritism, self-serving biases, value-action gaps, moral disengagement, etc. |
| **Fortitude** | Fortitude means idealism or the will to stick to moral ideals and moral responsibilities, potentially against all odds. For the AI sector, this means that researchers and managers acquire the courage to speak up when they come across moral issues. This may sound obvious, but in light of powerful situational forces, peer influences or authorities, speaking up and truly acting in accordance to one's own convictions can become very difficult. Fortitude helps to overcome these difficulties. | Situational forces, peer influences, authorities, etc. |

*Table 2 - List of second-order AI virtues*

In the following, a selection of some of the major factors of bounded ethicality that can be tackled by prudence shall be described. This selection is neither exhaustive nor does it go into much detail. However, it is meant to be a practical overview that can set the scene for more in-depth subsequent analyses.

Clearly, the most obvious factors of bounded ethicality are psychological biases (Cain and Detsky 2008). It is common that people's first and often only reaction to moral problems is emotional. Or, in other words, taking up dual-process theory, their reaction follows *system 1 thinking* (Kahneman 2012; Tversky and Kahneman 1974), meaning an intuitive, implicit, effortless, automatic mode of mental information processing. System 1 thinking predominates everyday decisions. System 2, on the other hand, is a conscious, logical, less error-prone, but slow and effortful mode of thinking. Although many decision making routines would require system 2 thinking, individuals often lack the energy to switch from system 1 to system 2. Ethical decision making needs cognitive energy (Mead et al. 2009). This is why prudence is such an important virtue, since it helps AI practitioners to transition from system 1 to system 2 thinking in ethical problems. This is not to say that the dual-process theory is without criticism. Recently, cognitive scientists have challenged its validity (Grayot 2020), even though they did not abandon it in toto. It still remains a scientifically sound heuristic in moral psychology. Thus, system 2 thinking remains strikingly close to critical ethical thinking, although it does obviously not necessarily result in it (Bonnefon 2018).

The transition from system 1 to system 2 thinking in ethical problems can also be useful for mitigating another powerful psychological force, namely *implicit biases* (Banaji and Greenwald 2013), that can impair at least two basic AI virtues, namely justice and care. Individuals have implicit associations, also called "ordinary prejudices", that lead them to classify, categorize and perceive their social surrounding with accordance to prejudices and stereotypes. This effect is so strong that even individuals who are absolutely sure to not be hostile towards minority groups actually are exactly that. The reason for that lies in the fact that people succumb to subconscious biases that reflect culturally established stereotypes or discrimination patterns. Hence, unintentional discrimination cannot be unlearned without changing culture, the media, the extent of exposure to people from minorities and the like. Evidently, this task cannot be fulfilled by the AI sector. Nevertheless, implicit biases can be tackled by increasing workforce diversity in AI firms and by using prudence as a virtue to accept the irrefutable existence and problematic nature of implicit biases as well as their influence on justice in the first place.

Another important bias that can compromise basic AI virtues and that can at the same time be overcome by prudence is *in-group favoritism* (Efferson et al. 2008). This bias causes people to sympathize with others who share their culture, organization, gender, skin color etc. For AI practitioners, this means that AI applications which have negative side-effects on outgroups, for instance the livelihoods of clickworkers in South-east Asia (Graham et al. 2017), are rated less ethically problematic than AI applications that would have similar consequences for in-groups. Moreover, the current gender imbalance in the AI field might be prolonged by in-group favoritism in human resource management. In-group favoritism mainly stifles character dispositions like justice and care. Prudence, on the other hand, is apt to work against in-group favoritism by recognizing artificial group constructions as well as definitions of who counts as "we" and who as "others", bolstering not only fair decision making, but also abilities to empathize with "distant" individuals.

One further and important effect of bounded ethicality that can impair the realization of the basic AI virtues are *self-serving biases*. These biases cause revisionist impulses in humans, helping to downplay or deny past unethical actions while memorizing ethical ones, resulting in a self-concept that depicts oneself as ethical.

When one asks individuals to rate how ethical they think they are on a scale of 0 to 100 related to other individuals, the majority of them will give themselves a score of more than 50 (Epley and Dunning 2000). The same holds true when people are asked to assess the organization they are a part of in relation to other organizations. Average scores are higher than 50, although actually the average score would have to be 50. What one can learn from this is that generally speaking, people overestimate their ethicality. Moreover, self-serving biases cause people to blame other people when things go wrong, but to view successes as being one's own achievement. Others are to blame for ethical problems, depicting the problems as being outside of one's own control. In the AI sector, self-serving biases can come into play when attributing errors or inaccuracies in applications as being the result of others, when reacting dismissive to critical feedback or feelings of concern, etc. Moreover, not overcoming self-serving biases by prudence can mean to act unjustly and dishonestly, further compromising basic AI virtues.

*Value-action-gaps* are another effect of bounded ethicality revealed by empirical studies in moral psychology (Godin et al. 2005; Jansen and Glinow 1985). Value-action gaps occur in the discrepancy between people's self-concepts or moral values and their actual behavior. In short, the gaps mark the distance between what people say and what people do. Prudence, on the other hand, can help to identify that distance. In the AI field, value-action gaps can occur on an organizational level, for instance by using lots of ethics-related terms in corporate reports and press releases while actually being involved in unethical businesses practices, lawsuits, fraud etc. (Loughran et al. 2009). Especially the AI sector is often accused of ethics-washing, hence of talking much about ethics, but not acting accordingly (Hao 2019). Likewise, value-action gaps can occur on an individual level, for instance by holding AI safety or data security issues in high esteem while actually accepting improper quality assurance or rushed development and therefore provoking technical vulnerabilities in machine learning models. Akin to value-action gaps are behavioral forecasting errors (Diekmann et al. 2003). Here, people tend to believe that they will act ethically in a given situation *X*, while when situation *X* actually occurs, they do not behave accordingly (Woodzicka and LaFrance 2001). They underestimate the extent to which they will indeed stick to their ideals and intentions. All these effects can interfere negatively with basic AI virtues, mostly with care, honesty and justice. This is why prudence with regard to value-action gaps is of great importance.

The concept of *moral disengagement* is another important factor in bounded ethical decision making (Bandura 1999). Techniques of moral disengagement allow individuals to selectively turn their moral concerns on and off. In many day-to-day decisions, people act contrary to their own ethical standards, but without feeling bad about it or having a guilty conscience. The main techniques in moral disengagement processes comprise justifications, where wrongdoing is justified as means to a higher end; changes in one's definition about what is ethical; euphemistic labels, where individuals detach themselves from problematic action contexts by using linguistic distancing mechanisms; denial of being personally responsible for particular outcomes, where responsibility is attributed to a larger group of people; the use of comparisons, where own wrongdoings are relativized by pointing at other contexts of wrongdoings or the avoidance of certain information that refers to negative consequences of one's own behavior. Again, prudence can help to identify cases of moral disengagement in the AI field and act as a response to it. Addressing moral disengagement with prudence can be a requirement to live up to all basic AI virtues.

In the following, a selection of some of the major factors of bounded ethicality that can be tackled by fortitude shall be described. Here, supra-individual issues that can impair ethical decision making in AI research and development are addressed. Certainly, one of the most relevant factors one has to discuss in this context are *situational forces* (Haney et al. 1973). Numerous empirical studies in moral psychology have shown that situational forces can have a massive impact on moral behavior (Williams and Bargh 2008; Isen and Levin 1972; Latané and Darley 1968). Situational forces can range from specific influences like the noise of a lawnmower that significantly affects helping behavior (Mathews and Canon 1975) to more relevant factors like competitive orientations, time constraints, tiredness, stress etc., which are likely to alter or overwrite ethical concerns (Cave and ÓhÉigeartaigh 2018; Darley and Batson 1973; Kouchaki and Smith 2014). Especially financial incentives have a significant influence on ethical behavior. In environments that are structured by economic imperatives, decisions that clearly have an ethical dimension can be reframed as pure business decisions. All in all, money has manifold detrimental consequences for decision making since it leads to decisions that are proven to be less social, less ethical or less cooperative (Vohs et al. 2006; Palazzo et al. 2012; Kouchaki et al. 2013; Gino and Mogilner 2014; Gino and Pierce 2009). Ultimately, various finance law obligations or monetary factual constraints that a company's management has to comply to can conflict with or overwrite AI virtues. Especially in contexts like this, virtue ethics can significantly be pushed into the background, although the perceived constraints lead to immoral outcomes. In short, situational forces can have negative impacts on unfolding all four basic AI virtues, namely justice, honesty, responsibility and care. Fortitude, on the other hand, is supposed to help to counteract these impacts, allowing the mentioned virtues to flourish.

Similar to and often not clearly distinguishable from situational forces are *peer influences* (Asch 1951; Asch 1956). Individuals want to follow the crowd, adapt their behavior to that of their peers and act similarly to them. This is also called conformity bias. Conformity biases can become a problem for two reasons: First, group norms can possess unethical traits, leading for instance to a collective acceptance of harm. Second, the reliance on group norms and the associated effects of conformity bias induces a suppression of own ethical judgements. In other words, if one individual starts to misbehave, for instance by cheating, others follow suit (Gino et al. 2009). A similar problem occurs with *authorities* (Milgram 1963). Humans have an inert tendency for being obedient to authorities. This willingness to please authorities can have positive consequences when executives act ethically themselves. If this is not the case, the opposite becomes true. For AI ethics, this means that social norms that tacitly emerge from AI practitioner's behavioral routines as well as managerial decisions can both bolster ethical as well as unethical working cultures. In case of the latter, the decisive factor is the way individuals respond to inner normative conflicts with their surroundings. Do they act in conformity and obedience even if it means to violate basic AI virtues? Or do they stick to their dispositions and deviate from detrimental social norms or orders? Fortitude, one of the two second-order virtues, can ensure the appropriate mental strength to stick to the right intentions and behavior, be it in cases where everyone disobeys a certain law but oneself does not want to join in, where managerial orders instruct to bring a risky product to the market as fast as possible but oneself insists on piloting it before release or where under extreme time pressure one insists on devoting time to understand and analyze training data sets.

## 5 Ethics training – AI virtues come into being

Cultivating basic and second-order AI virtues means achieving virtuous practice embedded in a specific organizational and cultural context. A virtuous practice requires some sort of moral self-cultivation that encompasses the acquirement of motivations or the will to take action, knowledge on ethical issues, skills to identify them and moral reasoning to make the right moral decisions (Johnson 2017). One could reckon that especially aforementioned skills or motivations are either innate or the result of childhood education. But ethical dispositions can be changed by education in all stages of life, for instance by powerful experiences or a certain work atmosphere in organizations. To put it in a nutshell, virtues can be trained and taught in order to foster ethical decision making and to overcome bounded ethicality.

The simplest form of ethics programs comprise ethics training sessions combined with incentive schemes for members of a given organization that reward the abidance of ethical principles and punish their violation. These ethics programs have numerous disadvantages. First, individuals that are part of them are likely to only seek to perform well on behavior covered by exactly these programs. Areas that are not covered are neglected. That way, ethics programs can even increase unethical behavior by actually well-intended sanctioning systems (Gneezy and Rustichini 2000). For instance, in case a fine is put on a specific unethical behavior, individuals who benefit from this behavior might simply weigh the advantage of the unethical behavior against the disadvantage of the fine. If the former outweighs the latter, the unethical behavior might even increase if a sanctioning system is in place. Ethical decisions would simply be reframed as monetary decisions. In addition to that, individuals can become inclined to trick incentive schemes and reward systems. Moreover, those programs solely focus on extrinsic motivators and do not change intrinsic dispositions and moral attitudes. All in all, ethics programs that comprise simple reward and sanctioning systems – as well as corresponding surveillance and monitoring mechanisms – are very likely to fail.

A further risk of ethics programs or ethics training are reactance phenomena. Reactance occurs when individuals protest against constraints of their personal freedoms. As soon as ethical principles restrict the freedom of AI practitioners doing their work, they might react to this restriction by trying to reclaim that very freedom by all means (Dillard and Shen 2005; Hong 1992; Dowd et al. 1991). People want to escape restrictions, thus the moment when such restrictions are put in place – no matter whether they are justified from an ethical perspective or not – people might start striving to break free from them. Ultimately, "forcing" ethics programs on members of an organization is not a good idea. Ethics programs should not be decoupled from the inner mechanisms and routines of an organization. Hence, in order to avoid reactance and to fit ethics programs into actual structures and routines of an organization, it makes sense to carefully craft specific, unique compliance measures that take particular decision processes of AI practitioners and managers into account. In addition to that, ethics programs can be implemented in organizations with delay. This has the effect of a "future lock-in" (Rogers and Bazerman 2008), meaning that policies achieve more support, since the time delay allows for an elimination of the immediate costs of implementation, for individuals to prepare for the respective measures and for a recognition of their advantages.

Considering all of that, what measures can actually support AI practitioners and AI companies' managers to strengthen AI virtues? Here, again, insights from moral psychology as well as behavioral ethics research can be used (Hines et al. 1987; Kollmuss and Agyeman 2002; Treviño et al. 2006; Treviño et al. 2014) to catalogue measures that bolster ethical decision making as well as virtue acquisition (see table 3 and 4). The measures can be vaguely divided into those that tend to affect single individuals and those that bring about or relate to structural changes in organizations. The following table 3 lists measures that relate to AI professionals on an individual level.

| Measures related to individuals | Explanation |
|---|---|
| **Knowledge about AI virtues** | AI professionals must be familiar with the six AI virtues and know about their importance and implications. |
| **Knowledge about action strategies** | Professionals have to learn how they can mitigate ethically relevant problems, for instance in the fields of fairness, robustness, explainability, but also in terms of organizational diversity, clickwork outsourcing, sustainability goals etc. |
| **Locus of control** | Professionals should have the perception that they themselves are able to influence and have a tangible impact on ethically relevant issues. This also supports a sense of responsibility, meaning that professionals hold themselves accountable for the consequences of their decision making. |
| **Public commitment** | Professionals can explicitly communicate the willingness to take action in ethical challenges. Publicly committing to stick to particular virtues, ideals, intentions and moral resolutions causes individuals to feel strongly obliged to actually do so when encountering respective choices. |
| **Audits and discussion groups** | With the help of colleagues, one can reflect and discuss professional choices, ethical issues or other concerns in one's daily routines in order to receive critical feedback. Furthermore, fictious ethical scenarios simulating particular contexts of decision making that professionals may face can be used. Apart from scenario trainings, organizations can grant professionals time for contemplation, allowing time to read texts e. g. about moral psychology or ethical theory. |

*Table 3 – Individual measures that bolster ethical decision making and virtue acquisition*

The following table 4 lists systemic measures that affect organizations mainly on a structural level.

| Systemic measures | Explanation |
|---|---|
| **Leader influences** | Managers play a key role as role models influencing employee's attitudes and behaviors. Their decisions have a particular legitimacy and credibility, which makes employees imitating them very likely. This way, managers, whose prosocial attitudes, fairness and behavioral integrity are of utmost importance, can define ethical standards in their organizations, since their way of making moral decisions trickles down to subordinate individuals (Treviño et al. 2014). |
| **Ethical climate and culture** | Unlike ethics codes, which are proven to have no significant effect on (un)ethical decision making in organizations, ethical climates do have that effect (McNamara et al. 2018). Especially caring |
| | climates are positively related to ethical behavior. On the other hand, self-interested, egoistic climates are negatively associated with ethical choice. Furthermore, ethical cultures, meaning informal norms, language, rituals etc., also affect ethical decision making and can, among other things, be significantly influenced by performance management systems (Kish-Gephart et al. 2010). |
| **Proportion of women** | Countless studies in empirical business ethics research indicate that women are more sensitive to and less tolerant of unethical activities than their male counterparts (Loe et al. 2013). In short, gender is a predictor to ethical behavior. This points out the importance of raising the proportion of female employees. Especially in the AI sector, male researchers currently strikingly outnumber females. This lack of workforce diversity has consequences on the functionality of software applications as well as implications on ethical outcomes in AI organizations. Hence, raising the proportion of women in the AI sector should pose one of the most effective measures to improve ethical decision making on a grand scale. This is not to say that the same doesn't hold true for other underrepresented demographics or marginalized populations. Here, the paper only points at the hiring of women, though. This is due to the fact that only for women and not for other demographic groups, ample research shows that they are less likely to engage in unethical behavior compared to men. |
| **Decreasing stress and pressure** | Reducing the amount of stress and time pressure in organizations can have game-changing consequences for the organizations' ethical climates (Darley and Batson 1973; Selart and Johansen 2011). De-stressing professionals, slowing down processes, and by that, setting cognitive resources free, promotes a transition from system 1 to system 2 thinking in decision making situations. This way, simply speaking, individuals are encouraged to think before they act, which can ultimately improve ethicality in organizations. |
| **Openness for critique** | Critical voices from the public can point at blind spots or specific shortcomings of organizations. Being open to embrace external critique as an opportunity to reflect upon an organization's own routines and goals with the associated willingness to potentially realign them can significantly improve its own trustworthiness, reputation and public perception. Eventually, this can contribute to the overall success of an organization. |

*Table 4 - Systemic measures that bolster ethical decision making and virtue acquisition*

# 6   Discussion

Virtue ethics doesn't come without shortcomings. In general, it is criticized for focusing on the "being" rather than the "doing", meaning that virtue ethics is agent- and not act-centered. Moreover, critics fault that on the one hand, virtuous persons can perform wrong actions, and on the other hand, right actions can be performed by persons who are not virtuous. However, this is a truism that could easily be transferred to other approaches in ethical theory, for instance by pointing at the fact that normative rules can be disregarded or violated by individuals or that individuals can perform morally right actions without considering normative rules. Another response to that critique stresses that it is one of virtue ethics' major strength to not universally define "right" and "wrong" actions. Virtue ethics can address the question of "eudaimonia" without fixating axiological concepts of what is "right". Further, virtue ethics is criticized by pointing at its missing "codifiability". Stipulating sets of virtues is arbitrary. However, this critique also holds true for every other ethical theory. Their very foundations are always arbitrary. All in all, many points of criticism that are brought into position in order to find faults in virtue ethics can equally be brought into position against other ethical theories, such as deontology or consequentialist ethics.

Apart from that, modern ethical approaches to ensure AI applications to meet ethical requirements are participatory approaches. Here, especially citizen science projects or third-party audits come into play. Volunteer involvement means attaching importance to voices from outside of one's own institute or even outside of the sciences and industry. Not only research itself can profit from citizens and independent third-party inspections (Ceccaroni et al. 2019; Zicari 2020), since they can also reflect on the research's very own ethical dimensions. While virtue ethics, especially care, can strengthen AI practitioners' sense for the impact of their actions on others, participatory approaches can directly give others a voice and step into the breach when virtue ethics should fail or be implemented incompletely.

# 7   Conclusion

Hitherto, all the major AI ethics initiatives choose a principled approach. They stipulate a list of rules and standards, suggesting that this has an effect on AI research and development. But, as more and more papers from AI-metaethics show (Mittelstadt 2019; Lauer 2020; Hagendorff 2020c; Rességuier and Rodrigues 2020), this approach fails in many regards. The deontological approach in AI ethics has no reinforcement mechanisms, it is not sensitive to different contexts and situations, it fails to address the technical complexity of AI, it uses terms and concepts that are often too abstract to be put into practice etc. In order to improve the two last-named shortcomings, AI ethics recently underwent a practical turn, stressing its will to put principles into practice and to finally have a tangible effect. But the typologies and guidelines on how to put AI ethics into practice stick to the principled approach altogether (Morley et al. 2020; Hallensleben et al. 2020) – and with that, they stick to the problems of deontology. But another approach, namely virtue ethics, seems to be more promising.

The goal of this paper was to outline how AI ethics can actually be put into practice by focusing on specific AI virtues. Virtue ethics focusses on an individual's character development. Character dispositions provide the basis for professional decision making. On the one hand, the paper considered insights from moral psychology on the many pitfalls the motivation of moral behavior has. On the other hand, it used virtue instead of deontological ethics to promote and foster not only four basic AI virtues, but also two second-order AI virtues that can help to circumvent "bounded ethicality" and one's vulnerability to unconscious biases. The basic AI virtues comprise justice, honesty, responsibility and care. Each of these virtues motivates a kind of professional decision making that builds the bedrock for fulfilling all the AI specific ethics principles discussed in literature. In addition to that, the second-order AI virtues, namely prudence and fortitude, can be used to overcome specific effects of bounded ethicality that can stand in the way of the basic AI virtues, meaning biases, value-action gaps, moral disengagement, situational forces, peer influences and the like. Lastly, the paper described framework conditions and organizational measurements that can help to realize ethical decision making and virtue training in the AI field. Equipped with this information, organizations dealing with AI research and development can actually put AI ethics into practice.

## Acknowledgements

This research was supported by the Cluster of Excellence "Machine Learning – New Perspectives for Science" funded by the Deutsche Forschungsgemeinschaft (DFG, German Research Foundation) under Germany's Excellence Strategy – Reference Number EXC 2064/1 – Project ID 390727645.

## Publication bibliography